\documentclass[11pt]{article}

\usepackage[dvips,letterpaper]{geometry}

\usepackage{srcltx}

\usepackage{amsmath}
\usepackage{amssymb,amsfonts}

\usepackage{epsfig}
\usepackage[usenames,dvipsnames]{color}
\usepackage[usenames,dvipsnames]{xcolor}
\usepackage{subfigure}

\usepackage{booktabs}

\usepackage{fullpage}
\usepackage{setspace}
\usepackage{flushend}
\usepackage{multicol}

\usepackage{url}

\usepackage[short,12hr]{datetime}
\usepackage{siunitx}

\usepackage{hyperref}

\makeatletter

\newenvironment{rsmallmatrix}{\null\,\vcenter\bgroup
  \Let@\restore@math@cr\default@tag
  \baselineskip6\ex@ \lineskip1.5\ex@ \lineskiplimit\lineskip
  \ialign\bgroup\hfil$\m@th\scriptstyle##$&&\thickspace\hfil
  $\m@th\scriptstyle##$\crcr
}{%
  \crcr\egroup\egroup\,%
}
\makeatother

\title{Multi-Beam RF Aperture \\ Using Multiplierless FFT Approximation}

\author{
Dora~Suarez$^\ast$
\quad
Renato~J.~Cintra%
\thanks{%
D. Suares and R. J. Cintra are with the
Signal Processing Group,
Departamento de Estat\'{\i}stica, 
Universidade Federal de Pernambuco.
E-mail: \protect\url{rjdsc@dsp.ufpe.org}}
\quad
F\'abio~M.~Bayer%
\thanks{%
F. M. Bayer is with the
Departamento de Estat\'{\i}stica and LACESM, 
Universidade Federal de Santa Maria, RS, Brazil.}
\\
Arindam Sengupta$^\ddagger$
\quad
Sunera Kulasekera$^\ddagger$
\quad
Arjuna Madanayake%
\thanks{%
A. Sengupta, S. Kulasekera and A. Madanayake
are with the
ECE, The University of Akron, Akron, OH, USA.}
}

\date{\today\ @ \currenttime}

\begin{document}

\maketitle

\onehalfspacing

\begin{abstract}
Multiple independent radio frequency (RF) beams find applications in communications, radio astronomy, radar, and microwave imaging.
An $N$-point FFT applied spatially across an array of receiver antennas provides $N$-independent RF beams
at $\frac{N}{2}\log_2N$ multiplier complexity. 
Here, 
a low-complexity multiplierless approximation for the 8-point FFT is presented for RF beamforming, using only 26~additions.
The algorithm provides eight beams that closely resemble the antenna array patterns of the traditional FFT-based beamformer albeit without using multipliers.
The proposed FFT-like algorithm is useful for low-power RF multi-beam receivers;
being
synthesized in 45~nm CMOS technology at 1.1~V supply, and verified on-chip using a
Xilinx Virtex-6 Lx240T FPGA device.
The CMOS simulation and FPGA implementation indicate bandwidths of 588~MHz and 369~MHz, respectively, for each of the independent receive-mode RF beams.
\end{abstract}

\maketitle

\section{Introduction}

Antenna array based radio frequency (RF) applications such as radar, wireless communications, localization, remote sensing, signal intelligence, radio astronomy, 
search for extraterrestrial intelligence (SETI),
and  imaging requires the fundamental operation of receive mode beamforming. 
To wit, 
beamforming is precisely the directional enhancement of propagating electromagnetic planar-waves based on their directions of arrival (DOA), whilst suppressing undesired noise and interference that impinge on the antenna array.
The ability to form multiple receiver beams is known as ``multi-beamforming''~\cite{Ellingson,ONR}.
Multiple RF beams, each having a unique ``look direction''---the direction of maximum sensitivity---is needed for multiple visibilities.

\begin{figure}%
\centering
\scalebox{0.9}{\input{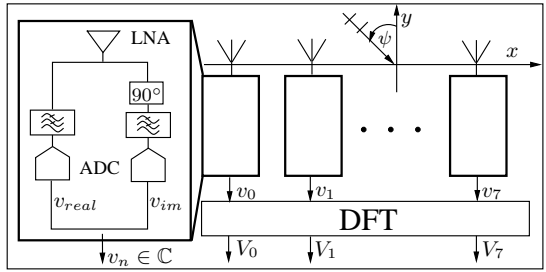}}
\caption{ULA-based multi-beamformer using a spatial FFT.}
\label{figure-dft-block}
\end{figure}

Multiple simultaneous beams are also needed for search-and-track radar, which in volume-scan mode, continuously monitor airborne threats, such as aircraft, warheads and cruise missiles, across a given range of angles. From the standpoint of high-capacity wireless communications, simultaneous receiver beams are of importance to multi-input multi-output (MIMO) systems.
The application of an 
$N$-point fast Fourier transform (FFT)---at each time sample---spatially, across a uniform linear array (ULA) of antennas, is a  technique for achieving a plurality of independent RF beams\cite{Ellingson,ONR}.
The FFT efficiently computes 
the discrete Fourier transform (DFT) with $\frac{N}{2}\log_2N$
multiplications.
Fig.~\ref{figure-dft-block} shows an overview of a ULA-based multi-beamformer using a spatial FFT. 
For an $N$-element ULA, the spatial FFT beamformer provides
$N$ beams, each uniformly spaced in the frequency domain
by the interval $2\pi/N$. The signal is first sent through a low noise amplifier (LNA) 
and 
the real
(I, in-phase, $v_{real}$) 
and 
the imaginary 
(Q, quadrature, $v_{im}$) parts are 
low-pass filtered and sampled using
analog-to-digital converters (ADCs), 
before application of the DFT.
The spatial angle $\psi$
is the independent variable used in the polar array beam-patterns.

RF aperture power consumption is directly proportional to circuit complexity and clock frequency.
Because multiplier hardware dominates circuit complexity, the utilization of FFT hardware having as small a number of parallel multiplier circuits as possible is preferable in terms of reduction of overall circuit complexity and power consumption of the multi-beamformer.
The proposed fast algorithm approximates the FFT computation without using any multipliers at all, making the corresponding digital architecture very simple to realize on-chip.
Because the proposed fast algorithm only requires 26 addition operation, the corresponding architecture is of lower power consumption compared to usual FFT-based circuits having parallel multipliers to implement the \emph{twiddle factors}.

\section{Multiplier-Free DFT Approximation}

The DFT is a linear orthogonal transformation relating 
an $N$-point 
input vector 
$\mathbf{v}= \begin{bmatrix}v_{0} & v_1 & \ldots &v_{N}\end{bmatrix}^\top$
to
an output vector
denoted by
$\mathbf{V}=\begin{bmatrix}V_{0} & V_1 & \ldots & V_{N} \end{bmatrix}^\top$
according to
$V_{k}
=
\sum_{n=0}^{N-1}
v_{n}
\cdot
\omega_N^{kn}
,
\quad 
k=0,1,\ldots,N-1$,
where 
$\omega_N = \exp\left\{-2\pi j/N\right\}$
is the $N$th root of unity~\cite{Oppenheim2009}
and
$j=\sqrt{-1}$. 
In matrix formalism,
the above expression reduces to:
$\mathbf{V} = \mathbf{F}_{N} \cdot \mathbf{v}$,
where $\mathbf{F}_{N}$ 
is the DFT matrix,
whose 
$(i,k)$-th element is given by
$f_{i,k}=\omega_N^{ik}$,
for $i,k=0,1,\ldots,N-1$.
The direct DFT computation
requires
$N^{2}$ complex multiplications and $N\cdot(N-1)$ additions.
Thus,
fast algorithms are necessary and are often able
to reduce the computation cost of the DFT computation
to $\mathcal{O}(N\cdot\log_2N)$ multiplications~\cite{Blahut2010}.

We submitted
the 8-point DFT matrix~$\mathbf{F}_8$
to the parametric-based optimization method
described in~\cite{Potluri2012}
to
derive
a matrix approximation.
Two major constraints were imposed on the sought approximations:
(i)~near-orthogonality 
and
(ii)~low-complexity.
Thus,
we obtained that the optimal
elements
for
the parametric approximation
of $\mathbf{F}_8$ are
$1$,
$(1-j)/2$,
and
$-j$.
Such parameters result
in the following
matrix approximation:
\begin{align*}
\hat{\mathbf{F}}_8
= 
\frac{1}{2}
\cdot
\left[
\begin{rsmallmatrix}
  2 & 2 & 2 & 2 & 2 & 2 & 2 & 2 \\ 
  2 & 1-j & -2j & -1-j & -2 & -1+j & 2j & 1+j \\ 
  2 & -2j & -2 & 2j & 2 & -2j & -2 & 2j \\ 
  2 & -1-j & 2j & 1-j & -2 & 1+j & -2j & -1+j \\ 
  2 & -2 & 2 & -2 & 2 & -2 & 2 & -2 \\ 
  2 & -1+j & -2j & 1+j & -2 & 1-j & 2j & -1-j \\ 
  2 & 2j & -2 & -2j & 2 & 2j & -2 & -2j \\ 
  2 & 1+j & 2j & -1+j & -2 & -1-j & -2j & 1-j \\
\end{rsmallmatrix}
\right]
.
\end{align*}
Compared to the exact DFT matrix,
above approximation has a mean squared error of
$0.686$,
which is considered low.
Although not exactly orthogonal,
the proposed approximation is very close to orthogonality.
Considering the deviation from orthogonality measure~\cite{Flury1986},
the proposed transform displayed a deviation of 0.03;
whereas,
in comparison,
the popular non-orthogonal DCT approximation SDCT~\cite{Haweel2001}
has a deviation from orthogonality of 0.20. 

\begin{figure}
\centering
\includegraphics[scale=0.95]{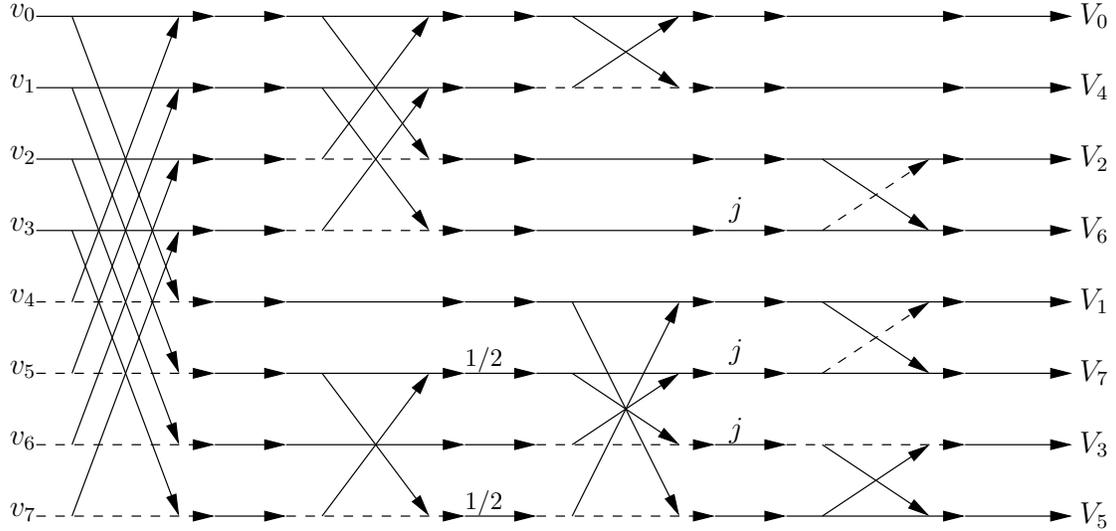}
\caption{Signal flow graph for the factorization of
$\hat{\mathbf{F}}_8$.
Input data 
$v_i$, 
$i=0,1,\ldots,7$, 
relates to the output
$V_k$, 
$k=0,1,\ldots,7$.
Dotted arrows
represent multiplications by
$-1$.}
\label{fig:flow}
\end{figure}

The proposed approximate matrix~$\hat{\mathbf{F}}_8$ 
preserves the symmetry of the DFT
and has null multiplicative complexity.
Still requiring 64~additions 
and 32 bit-shifting operations,
a further reduction in the additive complexity
can be obtained by means of a tailored fast algorithm.
Let $\mathbf{I}_n$ 
be
the
identity matrix of order $n$
and
$\mathbf{B}_n = 
\left[\begin{rsmallmatrix}1 & 1\\ 1 & -1 \end{rsmallmatrix}\right]
\otimes
\mathbf{I}_{n/2}$,
where $\otimes$ denotes the Kronecker product.
Thus,
employing the matrix factorization methods
suggested in~\cite{Blahut2010},
we
have the following fast algorithm:
\begin{align*}
\mathbf{\hat{F}}_{8}
=&
\mathbf{P}
\times
\operatorname{diag}
\big(
\mathbf{I}_2,\mathbf{A}_1,\mathbf{A}_3
\big)
\times
\mathbf{D}_2
\times
\operatorname{diag}
\big(
\mathbf{B}_2,\mathbf{I}_2,\mathbf{A}_4
\big)\\
&
\times
\mathbf{D}_1
\times
\operatorname{diag}
\big(
\mathbf{B}_{4},\mathbf{A}_2
\big)
\times
\mathbf{B}_8
,
\end{align*}
where
$
\mathbf{A}_{1}=
\left[
\begin{rsmallmatrix}
1	&	-1\\
1	&	1
\end{rsmallmatrix}
\right]
$,
$
\mathbf{A}_{2}=
\left[
\begin{rsmallmatrix}
1 & & & \\
  & 1	&	& 1\\
  & &	1 & \\
& 1 & & -1
\end{rsmallmatrix}
\right]
$,
$
\mathbf{A}_{3}=
\left[
\begin{rsmallmatrix}
1	&-1	& & \\
&	 & -1 & 1 \\
1 & 1 &  & \\
 & & 1& 1
\end{rsmallmatrix}
\right]
$,
$
\mathbf{A}_{4}=
\left[
\begin{rsmallmatrix}
1	&	& &1 \\
&	1 & 1 & \\
&1 & -1 &\\
1 & & & -1
\end{rsmallmatrix}
\right]
$,
$\mathbf{D}_1=\operatorname{diag}
(
\begin{rsmallmatrix}
1, & 1, & 1, & 1, & 1, & 1/2, & 1, & 1/2
\end{rsmallmatrix}
)$,
$\mathbf{D}_2=\operatorname{diag}
(
\begin{rsmallmatrix}
1, & 1, & 1, & j, & 1, & j, & j, & 1
\end{rsmallmatrix}
)$,
$
\mathbf{P}
=
\left[
\begin{rsmallmatrix}
\mathbf{e}_1&\big|\,
\mathbf{e}_5&\big|\,
\mathbf{e}_3&\big|\,
\mathbf{e}_6&\big|\,
\mathbf{e}_2&\big|\,
\mathbf{e}_8&\big|\,
\mathbf{e}_4&\big|\,
\mathbf{e}_7
\end{rsmallmatrix}
\right]
^\top
$
is a
permutation matrix,
and
$\mathbf{e}_i$ is
the 8-point column vector with
element~1 at the $i$th position
and 0~elsewhere. 
Figure~\ref{fig:flow}
depicts the signal flow graph
of the introduced algorithm.
The arithmetic complexity assessment in
terms of real operations
and comparisons
are
summarized in Table~\ref{table:complexity}.

\begin{table}

\centering

\caption{Real operation assessment and comparison}
\label{table:complexity}

\begin{tabular}{lccc}
\toprule
Method & Multiplications & Additions & Shifts
\\
\midrule
Exact DFT  & 256 & 240 & 0
\\
\midrule
FFT (complex input)~\cite{Oppenheim2009} & 4 & 52 & 0
\\
\midrule
FFT (real input)~\cite{Oppenheim2009} & 2 & 26 & 0
\\
\midrule
Proposed (complex input) & 0 & 52 & 4
\\
\midrule
Proposed (real input) & 0 & 26 & 2
\\
\bottomrule
\end{tabular}
\end{table}

\begin{figure}
\centering
\subfigure[DFT-based]{\includegraphics[width=0.3\linewidth]{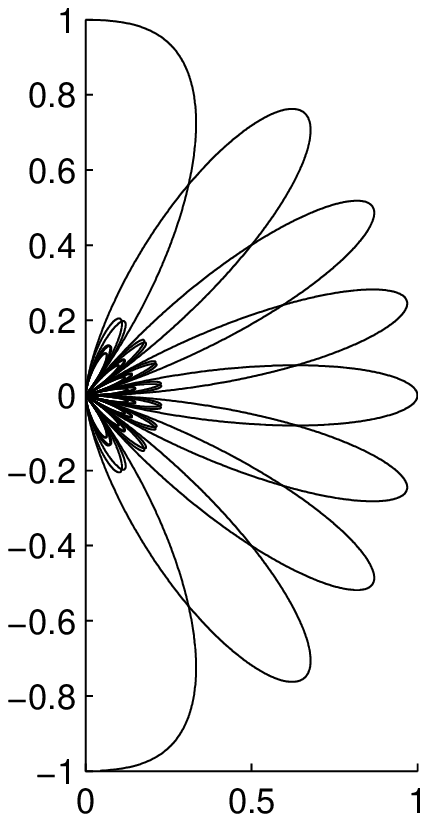}}
\quad
\subfigure[Proposed]{\includegraphics[width=0.3\linewidth]{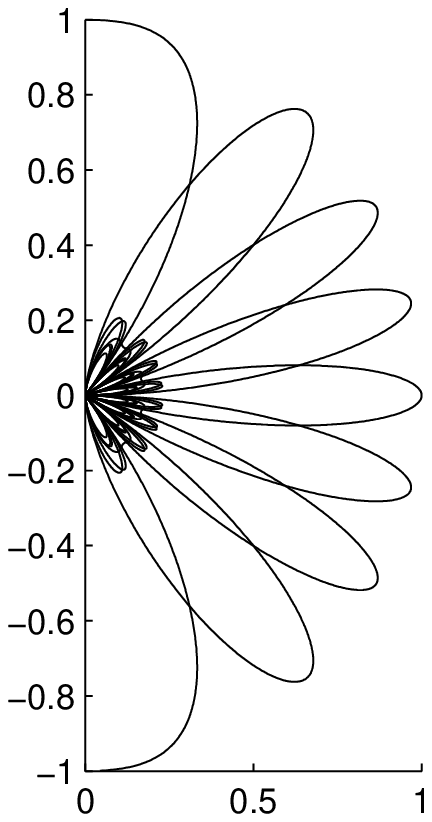}}
\quad
\subfigure[Error]{\includegraphics[width=0.3\linewidth]{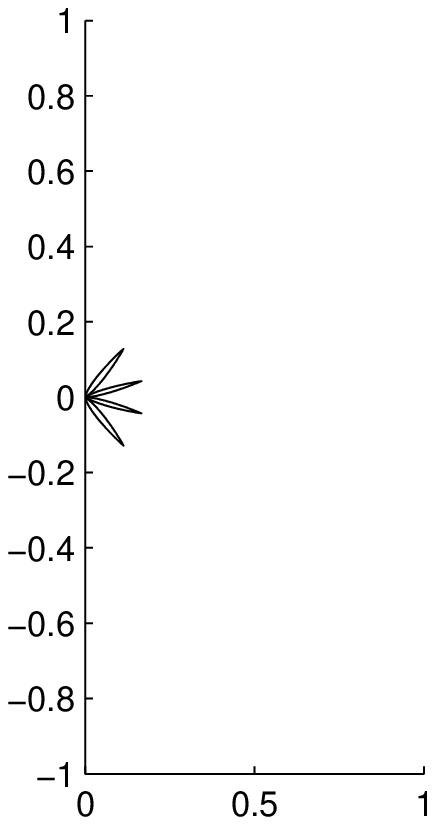}}
\caption{Polar plots of $P_i(\psi; \mathbf{F}_{8})$, 
$i=0,1,\ldots, 7$, 
$\psi\in[-\pi/2,\pi/2]$
at the frequency $\omega_t=\pi$ for the 
(a)~exact transform~$\mathbf{F}_8$,
(b)~proposed approximate transform~$\hat{\mathbf{F}}_8$, 
and 
(c)~error measure $D_i (\psi)$.}
\label{fig:transfer}
\end{figure}

Each row $i$ of matrix~$\mathbf{F}_8$
may be interpreted as 
the coefficients of
a
discrete filter
whose
transfer function is
$H_i (\omega; \mathbf{F}_{8}) 
= 
\sum_{k=0}^{7} 
f_{i,k} 
\cdot 
{\rm exp} (-jk \omega)$,
$i = 0,1, \ldots, 7$, 
for $\omega \in [-\pi, \pi]$~\cite{Oppenheim2009}.
In the case of multi-beam forming, 
the exact or approximate DFT
are applied spatially, 
across a ULA of antennas.
Here,
variable
$\omega$ is the \emph{spatial frequency} across the ULA. 
Let the normalized temporal frequency of the incident plane wave be 
$\omega_t\le\pi$.
From physics, 
we have that
$\omega=-\omega_t\sin\psi$, 
for
$-\pi/2\le\psi\le\pi/2$,
measured counter-clockwise from ULA broadside.
We set $\omega_t=\pi$,
which corresponds to $\psi \in [-\pi/2,\pi/2]$.
Thus,
the array patterns are given by:
\begin{align*}
P_i(\psi; \mathbf{F}_{8})
=
\frac{|H_i(-\omega_t \sin(\psi); \mathbf{F}_{8})|}
{\beta_i}
,
\end{align*}
where
$\beta_i = \max_\psi |H_i(-\omega_t \sin(\psi))|$,
for $i=0,1,\ldots,7$,
is a normalization factor.
\emph{Mutatis mutandis},
the array patterns 
based on the proposed approximation
are denoted by
$P_i (\psi, \hat{\mathbf{F}}_{8})$,
$i=0,1,\ldots,7$.
Figure~\ref{fig:transfer}(a)--(b) shows the
pattern arrays
associated to each
row of $\mathbf{F}_8$ and
$\hat{\mathbf{F}}_8$.
The eight independent beams are pointed at angles
$\psi_k = 0.00, \pm 14.47, \pm 30.00, \pm 48.59, 90.00$
in degrees
measured from array broadside direction, 
as expected from the conventional DFT beamformer.
To quantify the
difference between
corresponding
array patterns,
we considered the following 
error function:
\begin{align*}
D_i (\psi) 
\triangleq 
\Big | 
P_i (\psi; \mathbf{F}_{8})
-
P_i (\psi; \hat{\mathbf{F}}_{8})
\Big |
, 
\quad 
i = 0, 1, \ldots, 7
.
\end{align*}

In Figure~\ref{fig:transfer}(c),
the polar plot of 
$D_i(\psi)$
for all rows of $\hat{\mathbf{F}}_8$
is displayed.
The error energy can be obtained integrating
$D_i(\psi)$:
\begin{align*}
\epsilon_i
=
\int_{-\pi/2}^{\pi/2}
D_i^2 (\psi)
\operatorname{d}\psi
,
\quad
i=0,1,\ldots,7
.
\end{align*}
This computation furnished
$\epsilon_i = 1.08$, for odd $i$,
and
$\epsilon_i = 0$, for even $i$.
The total error energy is $4.32$.
For comparison,
the approximate DCT described in~\cite{Bouguezel2008}
has a total error energy of $4.12$.

\section{FPGA Realization and ASIC Synthesis}

The proposed multiplierless architecture was realized on digital hardware using an ML-605 Xilinx Virtex-6 field programmable gate array (FPGA) prototyping board.
The design was built and tested for 16-bit inputs via JTAG interface.
Moreover, 
it was pipelined to minimize the critical-path delay ($T_\text{cpd}$), which in turn offers the maximum frequency of operation and RF bandwidth.
The on-FPGA measured results verified the performance of the proposed architecture.
The FPGA resource consumption, 
including the number of slices, 
look-up tables (LUTs), and flip-flop (FF) count, 
is presented in Table~\ref{FPGA}. 
The percentage utilization of the available resources is also shown.
The pipelined design offered a maximum frequency of 739~MHz corresponding to a maximum RF bandwidth of 369~MHz for each of the eight beams.

\begin{table}
\centering

\caption{FPGA resource consumption}
\label{FPGA}

\begin{tabular}{lc}

\toprule %
Resources & Proposed  \\
\midrule %
Slice Registers & 3064 (1\%)
\\
\midrule
Slice LUTs & 2044 (1\%)
\\
\midrule
Occupied Slices & 620 (1\%)
\\
\midrule
LUT-FF Pairs & 2335 (1\%)
\\
\midrule
Bonded IOBs & 2 (1\%)
\\
\midrule
T$_{cpd}$ (ns) & 1.353
\\
\midrule
Max. Frequency (MHz) & 739.09
\\
\bottomrule 
\end{tabular}
\end{table}

\begin{table}
\centering

\caption{ASIC synthesis results}
\label{ASIC}

\begin{tabular}{lc}

\toprule %
Resources & Proposed
\\
\midrule %
Area (mm$^2$) & 0.064
\\
\midrule
Dynamic Power (mW) & 94.18
\\
\midrule
Static Power (mW) & 0.41
\\
\midrule
Total Power (mW) & 94.59
\\
\midrule
$T_\text{cpd}$ (ns) & 0.85
\\
\midrule
Max. Frequency (GHz) & 1.176
\\
\midrule
AT ($\text{mm}^2\text{ns}$) & 0.054
\\
\midrule
$\text{AT}^2$($\text{mm}^2\text{ns}^2$) & 0.046
\\
\bottomrule 
\end{tabular}
\end{table}

The FPGA-based digital design was imported to Cadence RTL compiler
for application-specific integrated circuit (ASIC) synthesis
using 45~nm complementary metal oxide semiconductor (CMOS) technology, for an operating voltage of 1.1~V at 27$^\circ$C. 
Table~\ref{ASIC}
displays
the
area, 
power,
critical path delay,
and maximum frequency of operation, at synthesis stage.
The area-time (AT) and area-time$^2$ (AT$^2$) complexities 
are reported. The CMOS synthesis shows an increase in the maximum clock frequency when compared to its FPGA implementation. 

\section{Conclusion}

An 8-point multiplierless
DFT approximation requiring 26~additions
was proposed.
Applications in receive mode
RF multi-beamforming using a ULA of antennas include communication, radar, and radio astronomy. CMOS synthesis and FPGA implementations have indicated bandwidths of 588~MHz and 369~MHz, respectively. The approximation is
 suitable for eight digital RF-beams, at low power.
The DFT approximation allows FFT-like performance without multiplier hardware.

\section*{Acknowledgements}
We thank 
CNPq, FACEPE, FAPERGS, 
and
The College of Engineering at UA
for 
the partial financial support.

\end{document}